\newcommand{\nn}{\nonumber}

\newcommand{\nb}{\nabla}

       \newcommand{\Dc}{ {\mathcal{D}} }

       \newcommand{\Gc}{ {\mathcal{G}} }

       \newcommand{\Lc}{ {\mathcal{L}} }

       \newcommand{\Rc}{ {\mathcal{R}} }

       \newcommand{\bb}{\mathbf{b}}
       \newcommand{\bW}{\mathbf{W}}

       \newcommand{\bx}{\mathbf{x}}
       
       \newcommand{\bp}{\mathbf{p}}

\documentclass[superscriptaddress,aip,amsmath,amssymb,reprint,noshowpacs]{revtex4-1}

\usepackage[utf8]{inputenc}
\usepackage[T1]{fontenc}
\usepackage{amsmath}
\usepackage{mathtools}
\usepackage{amsfonts}
\usepackage{amsmath}
\usepackage{graphicx}
\usepackage{dcolumn}
\usepackage{bm}
\usepackage[export]{adjustbox}
\usepackage{wrapfig}
\usepackage{lipsum}
\usepackage[titletoc,toc]{appendix}
\usepackage[colorlinks]{hyperref}
\usepackage{color}
\DeclareUnicodeCharacter{00D7}{$\times$}

\begin{document}

\title{Neural network tokamak equilibria with incompressible flows}

\author{D. A. Kaltsas}
\email{d.kaltsas@uoi.gr}
\affiliation{
Department of Physics, University of Ioannina,\\ GR 451 10 Ioannina, Greece}
\affiliation{Department of Physics, International Hellenic University,\\ GR 654 04 Kavala, Greece
}
\author{G. N. Throumoulopoulos}%
\email{gthroum@uoi.gr}
\affiliation{
Department of Physics, University of Ioannina,\\ GR 451 10 Ioannina, Greece}


\begin{abstract}
We present several numerical solutions to a generalized Grad-Shafranov equation (GGSE), which governs axisymmetric plasma equilibria with incompressible flows of arbitrary direction, using fully connected, feed-forward, deep neural networks, also known as multi-layer perceptrons. Such artificial neural network (ANNs) are trained to approximate tokamak-relevant equilibria upon minimizing the GGSE mean squared residual in the plasma volume and the poloidal flux function on the plasma boundary. Solutions for the Solovev and the general linearizing ansatz for the free functions involved in the GGSE are obtained and benchmarked against known analytic solutions. We also construct a non-linear equilibrium incorporating characteristics relevant to the high confinement mode. In our numerical experiments it was observed that changing the radial distribution of the training points has a surprisingly small effect on the accuracy of the trained solution. In particular it is shown that localizing the training points at the plasma edge results in ANN solutions that describe quite accurately the entire magnetic configuration, thus demonstrating the interpolation capabilities of the ANNs.

%
\end{abstract}

\pacs{Valid PACS appear here}
\maketitle


\section{Introduction}
Equilibrium and stability of fusion plasmas are commonly studied in the context of ideal Magnetohydrodynamics (MHD); in the case of tokamaks, which are toroidal and nearly axisymmetric magnetic confinement devices, the description is simplified geometrically by the assumption of axial symmetry. Equilibrium studies in the context of axisymmetric MHD rely on the solution of the so-called Grad-Shafranov equation (GSE) \cite{Shafranov1958,Grad1958}, which is usually solved for static plasmas or for purely toroidal plasma flows. For  arbitrary compressible flows, a nonlinear coupling between the GSE and the Bernoulli equation emerges, hence elaborate equilibrium solvers (see, e.g., Ref. \onlinecite{Guazzotto2004}) are required to determine equilibrium solutions. Macroscopic sheared flows are important for plasma confinement since they are associated with the suppression of certain instabilities and the formation of transport barriers that reduce the radial transport leading to improved confinement modes (High confinement mode -- H-mode) in tokamaks \cite{Burrell1997}. Therefore, a flexible modelling of sheared flows of arbitrary direction, particularly in connection with the H-mode phenomenology is desirable. Such a flexible and simplified description is provided by introducing the assumption of flow incompressibility, that decouples the GSE and the Bernoulli equation and ensures that the former stays in the elliptic regime. This implies that the incompressibility assumption is limited to low Mach number flows. 

It is known that stationary MHD states of axisymmetric plasmas with incompressible flows satisfy the following generalized Grad-Shafranov equation (GGSE) \cite{Tasso1998,Simintzis2001}
\begin{eqnarray}
\Delta^{*}u+\frac{1}{2}\frac{d}{du}\left[\frac{X^{2}}{1-M_{p}^{2}}\right]
+r^{2}\frac{dP_{s}}{du}+\nn \\+\frac{r^{4}}{2}\frac{d}{du}
\left[\rho\left(\frac{d\Phi}{du}\right)^{2}\right]=0\,, \label{ggse}
\end{eqnarray}
where $\Delta^*=r^2\nabla\cdot(\nabla/r^2)$ and ($r,\phi, z$) are cylindrical coordinates with $z$ corresponding to the axis of symmetry. The function $u=u(r,z)$ that labels the magnetic surfaces is associated with the ordinary magnetic flux function  $\psi$ via the following integral transformation \cite{Morrison1986aps}
\begin{eqnarray}
u(\psi)=\int_{0}^{\psi}\left[1-M_p^2(g)\right]^{1/2}dg\,. \label{integral_transform}
\end{eqnarray}
Here, $M_p(u)$ is the poloidal Mach-Alfv\'en function, i.e. the poloidal fluid speed over the
poloidal  Alfv\'en speed. In Eq. \eqref{ggse},
 $X(u)$ relates to the toroidal magnetic
 field, $B_\phi=I/r$,  through $I=(X-r^2\sqrt{\varrho}M_p\Phi^\prime)/(1-M_p^{2})$; $\Phi(u)$ is the electrostatic potential and $\rho(u)$ the plasma density;
 for vanishing flow the surface function $P_s(u)$
  coincides with the pressure and Eq. (\ref{ggse}) reduces to the usual (static) GS equation.

For fixed boundary equilibria, Eq. \eqref{ggse} is accompanied by a Dirichlet boundary condition of the form $u|_{\partial\Dc}=u_b$, where $u_b$ is a fixed value and $\partial\Dc$ is the boundary of the domain $\Dc\subset \mathbb{R}^2$. In present-day tokamaks the plasma boundary is of a characteristic D shape, which is elongated in the vertical direction and usually has a lower X-point associated with the presence of a plasma divertor. A flexible and accurate approximation of the computational domain and its boundary, which might be up-down symmetric or asymmetric, is required for constructing accurate tokamak equilibria. These equilibrium states are important for understanding and improving plasma confinement since they are used  not only for studying force-balance but also stability, and transport phenomena. In two previous papers we derived analytic solutions in analytically prescribed domains for the Solovev \cite{Solovev1968} and the general linearizing ansatz
\cite{Kaltsas2014,Kaltsas2019c} deploying an effective method for boundary shaping \cite{Cerfon2010}. However, these solutions are limited in the linear regime and finding solutions for nonlinear choices of the free functions in general requires the deployment of numerical methods, which are usually based on finite difference or finite element discretizations of the computational domain. For solutions obtained by conventional numerical methods, a sophisticated discretization is often required so as the boundary to be approximated with sufficient accuracy. In addition, these are not continuous, closed form solutions, so their evaluation in between the discrete nodes requires interpolation and the analysis is grid-dependent. 

Motivated by the growing interest in neural network and machine learning techniques for solving differential equations, we apply these ideas in the case of the GGSE, demonstrating the capability of neural network solutions to describe efficiently tokamak equilibria with flows of arbitrary direction, solving Eq. \eqref{ggse} for fixed boundaries. The method is accurate and grid-independent since no sort of discretization for the domain $\Dc$ and derivative approximations are needed as in classical numerical methods. This is accomplished upon using automatic differentiation (AD) \cite{Baydin2018,Raissi2019} to compute the partial derivatives of the neural network solution with respect to the input coordinates. For the training process we use randomly sampled points, and impose the boundary conditions on training points that are generated by parametric equations, thus giving us a great flexibility in shaping the boundary as desired. The method works efficiently for various boundary shapes and also for nonlinear GGSEs providing closed form ANN solutions, which are continuous and differentiable. Therefore, it is a potential candidate for treating some of the pathologies of the standard analytic and numerical methods. 

Neural Network Grad-Shafranov equilibria have already been constructed in the past, e.g. in Ref. \onlinecite{vanMilligen1995} and in Ref. \onlinecite{Joung2019} where the latter work was concerned with the reconstruction of magnetic equilibria using measured magnetic signals. Neural network representations of the magnetic flux were also employed in Ref. \onlinecite{Tribaldos1997} for rapid recovery of the plasma topology.   In Ref. \onlinecite{vanMilligen1995} the authors employed a procedure quite similar to the method described in Ref. \onlinecite{Raissi2019}, which is essentially the approach utilized also in the present paper. In this work though, we solve the generalized GS equation \eqref{ggse} rather than the static GSE and in addition, we use Deep Neural Networks  (DNNs) in contrast to Ref. \onlinecite{vanMilligen1995} where Multilayer Perceptrons  (MLPs) with only one hidden layer were employed. Also, our approach allows for analytically defined domains while is completely mesh-free, since the training points are randomly sampled, and aims to highlight the interpolation capabilities of the ANN solutions. These remarkable interpolation properties are demonstrated upon considering both radially uniform distributions of points, uniform in the sense that the point density is approximately constant throughout the computational domain, and point distributions localized near the plasma edge. Remarkably, the solutions obtained by these two settings have very small relative errors, with maximum values in the central region of the domain where in the second setting the training points are sparse or not present at all. 

The rest of the paper is organized as follows: in Sec. \ref{sec_II} we present some basic notions and methods regarding the ANN solutions to Partial Differential Equations (PDEs), then in Sec. \ref{sec_III} we describe the methodology we implemented for this particular problem, in Sec. \ref{sec_IV} the main results obtained by our numerical experiments are presented, and in Sec. \ref{sec_V}  we summarize the conclusions and discuss possible future extensions.

\medskip
\section{Artificial Neural Networks and PDEs}
\label{sec_II}
The idea of using ANNs and in particular MLPs to represent solutions to PDEs is documented in the seminal work of Lagaris et. al. in Ref. \onlinecite{Lagaris1998} and is based on the universal approximation theorem \cite{Hornik1989} which guarantees that MLPs are universal function approximators.  Interestingly, this approach was exploited even earlier for the static Grad-Shafranov equation in Ref. \onlinecite{vanMilligen1995} utilizing MLPs with only one hidden layer. The neural network solutions are still approximate, however, unlike conventional numerical solutions, they are continuous and their output values can be recovered everywhere inside the computational domain, once the weights and biases of the neural network are determined. Over the last years  this  subject field has attracted an explosively growing interest owing to recent developments involving, among others, the so-called Physics Informed Neural Networks (PINNs) \cite{Raissi2019}. Due to this growing interest in using neural networks to approximate solutions to boundary and initial value problems, several software packages have been developed in the last couple of years, for example we are aware of DeepXDE, NeuroDiffEq, PyDEns and Nangs
packages \cite{Lu2021,Chen2020,Koryagin2019,Pedro2019}. The results of the present study though, are obtained by an implementation built from scratch so that we could easily adapt the code to the needs of our specific problem. Although more information on these methods, e.g. for PINNs, can be found, in a plethora of references, e.g. Refs. \onlinecite{Lagaris1998, Raissi2019, Blechschmidt2021}, for reasons of completeness we describe here some of the basic principles for solving differential equations using neural networks.

The central idea is to turn the PDE and the accompanying boundary conditions, into an optimization problem where we aim to minimize the residual of the differential equation and the residual of the boundary conditions. A boundary value problem, like the fixed-boundary MHD equilibrium with incompressible flows, described by Eq. \eqref{ggse} and Dirichlet boundary conditions for $u$, can be written in the form
\begin{eqnarray}
\Rc:=\Lc u-\Gc(\bx,u)=0\,, \quad in\; \Dc \subset \mathbb{R}^d \nn\\
u(\bx)-u_b=0\,, \quad on\; \partial\Dc\,, \label{BVP}
\end{eqnarray}
where $\bx\in \mathbb{R}^d$, $u:	\mathbb{R}^d\rightarrow \mathbb{R}$, $\Lc$ is a differential operator and $\Gc$ a generally nonlinear function. In the case of the GGSE, $d=2$, $\Lc=\Delta^*$ and $\Gc=\Gc(r,u)$ (see Eq. \eqref{ggse}). A neural network solution $u_{n}=u_{n}(\bx;\bp)$, which is not an exact solution, produces a non vanishing residual error when substituted in \eqref{BVP}. Here, $\bp$ represents a vector composed of the neural network parameters.  These errors in the continuous setting can be estimated by 
\begin{eqnarray}
 L_i(\bp) &=& \frac{1}{V_{\Dc}}\int_{\Dc} d^dx \,\big|\Rc(\bx, u_{n}(\bx;\bp))\big|^2\,, \\
 L_{b}(\bp) &=& \frac{1}{S_{\partial \Dc}}\int_{\partial \Dc} d^{d-1} x\,\big|u_{n}(\bx;\bp)-u_b\big|^2\,,  
 \label{losses_1}
\end{eqnarray}
To perform the neural network training, the two loss functions are approximated using Monte-Carlo integration using a set of randomly sampled points. These points serve as the training data that are fed to the neural network whose parameters, i.e., the weights and the biases, are adapted so as the two loss functions are minimized. Note that the errors in the discrete problem are reduced to the Mean Squared Errors (MSEs). The minimization is effected by some gradient-based optimization method, e.g., the Gradient Descent method, leveraging the backpropagation algorithm to compute the derivatives of the loss functions with respect to the network parameters. Usually a weighted sum of the two losses is minimized, i.e.
\begin{eqnarray}
 L(\bp)=\lambda_i L_{i}(\bp)+\lambda_{b} L_{b}(\bp)\,. \label{total_loss}
\end{eqnarray}
Having formed the loss function \eqref{total_loss}, the optimization procedure can be expressed mathematically as follows:
\begin{eqnarray}
 \underset{\bp}{min}( L(\bp) )\,.
\end{eqnarray}

\subsubsection{Neural Network Architecture}
An MLP is a feed-forward, fully connected neural network with one or more hidden layers. In such ANNs the information moves only in one direction from the input to the output layer. Each neuron is connected with the outputs of the neurons in the previous layer by linear functions, containing weights  and amplitudes, called biases. These are the learnable model parameters, which are adapted (learned) during the training process. The output of each neuron is filtered by a nonlinear function, the so-called activation function $f$, before is fed to the next layer of neurons. Overall, an MLP can be written as a composition of alternating linear and non-linear transformations
\begin{eqnarray}
&&\hspace{-5mm} u_{n}(\bx;\bp)= \bW^{(\ell+1)}\cdot f^{(\ell)}(\bW^{(\ell)}\cdot f^{(\ell-1)}(\cdots \nn\\
&& \hspace{5mm} f^{(1)}(\bW^{(1)}\cdot\bx+\bb^{(1)})\cdots)+\bb^{(\ell)})+\bb^{(\ell+1)}\,, \label{nn_structure}
\end{eqnarray}
where $\ell$ is the number of the hidden layers. Here, $\bW^{(j)}$, $\bb^{(j)}$, $j=0,...,\ell+1$ are the weight matrices and the bias vectors, respectively, whose entries can be conflated into the vector $\bp$ of model parameters:
\begin{eqnarray}
& \bp=\left(W_{ik}^{(j)},b_{i}^{(j)}\right)\,,\nn\\
&  1\leq j\leq \ell+1\,, \; 1\leq i\leq n_j\,, \; 1\leq k\leq n_{j-1}\,,
\end{eqnarray}
where $n_j$ is the number of neurons in each layer $j$.

In our specific problem, $\bx=(r,z)$ and $u:\mathbb{R}^2\rightarrow \mathbb{R}$, hence, the input layer of the MLP should have two neurons: one for the $r$ coordinate and one for the $z$ coordinate, while it has a single output neuron representing the value of $u$ at the specific point $(r,z)$. We experimented with various MLP architectures varying the number of  hidden layers and the number of neurons per layer and  although we did not perform a rigorous optimization of the MLP, we empirically concluded that in most cases the algorithm is satisfactorily fast and accurate for an architecture with 4 hidden layers and 64 neurons per layer.  This entails a total number of 12737 model parameters (weights and biases) that have to be learned. As it can be deduced by the various loss history plots in Sec. \ref{sec_IV}, the loss function for a batch of interior validation points follows closely the loss computed on the inner training points (i.e. the GGSE MSE), hence we have no indications of overfitting on the sparse inner training data (1024 points) with this particular architecture\footnote{A set of validation points can be constructed by dividing the original batch of inner points in two sub-batches. The one is then used for training the network and the other for assessing its performance}. Presumably, better optimized architectures can be found, depending of course on the choice of the free functions and the free parameters of the GGSE, the complexity of the boundary and the number of training points.

Regarding the activation function, various choices have been considered, including the $tanh(x)$, the sigmoid function $\sigma(x)$ and the recently proposed $swish$ function \cite{Ramachandran2017}: $swish(x)=x\sigma(\gamma x)$, i.e.,
\begin{eqnarray}
 swish(x)=\frac{x}{1-e^{- \gamma x}}\,.\label{swish}
\end{eqnarray}
For $\gamma=1$ Eq. \eqref{swish} is also known as the Sigmoid-weighted Linear Unit (SiLU) activation function. In our implementation we treated $\gamma$ as a learnable parameter initializing it to unity and then letting the optimizer to adjust its value in each iteration (epoch). The $swish$ function has similarities with the infamous Rectified Linear Unit (ReLU) function, e.g., is bounded below and unbounded above, but unlike ReLU, it is smooth and its derivative is not a step function, so it can be used to MLPs approximating smooth solutions $u$.

\subsubsection{Forward pass and Automatic Differentiation}
Acting with the elliptic Shafranov operator on the neural network solution $u_n$ requires the computation of partial derivatives of $u_n$ with respect to the coordinates $r,z$. PINNs use Automatic Differentiation (AD) \cite{Baydin2018}, that requires a forward evaluation, which is straightforward in view of \eqref{nn_structure} and then a backward pass to compute the partial derivatives, applying the chain rule to differentiate $u_n$ with respect to the inputs $\bx$. Thus, the compositional structure of the neural networks allows us to efficiently and accurately compute derivatives without the truncation errors that are unavoidably involved in numerical differentiation. The same method is used for evaluating the network's derivatives with respect to the parameters $\bp$, that are required for the gradient-based optimization.

\medskip
\section{Implementation}
\label{sec_III}
\subsection{Loss functions}
\label{III_A}
In our implementation we carried out automatic differentiation and optimization using the PyTorch library \cite{Paszke2019}. In PyTorch AD is an automated procedure provided by the  automatic differentiation engine \textbf{autograd}. We should note here that our approach has a lot of similarities with Ref. \onlinecite{Pedro2019}, essentially extending it to a mesh-free implementation and increasing the flexibility, e.g., regarding the choice of activation functions, the definition of the computational domain and the distribution of the training points. 

The loss function that is to be minimized consists of two terms, one that measures the PDE residual error and one penalty term for the imposition of the boundary condition. These terms are:
\begin{eqnarray}
L_{i} &=& \frac{1}{N_{i}}\sum_{j=1}^{N_{i}}\Big|\Rc(r_j,z_j,u_{n}(r_j,z_j;\bp))\Big|^2\,,\; (r_j,z_j)\in \Dc\,,\nn\\ \label{loss_inner}\\
L_{b} &=& \frac{1}{N_{b}}\sum_{j=1}^{N_{b}}\big|u_{n}(r_j,z_j;\bp)-u_b\big|^2\,, \;  (r_j,z_j)\in \partial \Dc\,, \label{loss_boundary}\\
L &=&\lambda_i L_{i}+\lambda_b L_{b}\,.
\end{eqnarray}
 In our runs we chose $N_i=N_b=N$ and we observed that the magnetic surfaces conform with the imposed boundary  for $\lambda_b > \lambda_i$;  otherwise, the boundary error could not be minimized sufficiently and as a consequence the computed boundary did not coincide with the imposed one. For the training process we applied two different approaches, the first was to update the learnable parameters twice in each epoch (iteration) first working through the boundary points and then through the inner points, which can be considered as a mini-batch gradient descend optimization with batch size equal to $N$ from a training set consisting of $2N$ points. The second approach was a full batch gradient descent optimization, backpropagating the error and updating the parameters once per epoch. Both implementations had similar outcomes in terms of training efficiency.

\subsection{Training data}
\label{III_B}
In this study the plasma domain and its boundary are parametrically prescribed upon using appropriate analytic formulas. We investigate two different cases, one with an up-down symmetric, smooth D-Shaped boundary and the case of up-down asymmetric boundary with a lower X-point topology. In this section we present the former case which is employed for benchmarking the neural network solutions against analytic ones. For the up-down symmetric configuration the computational domain can be defined by the following parametric equations \citep{Turnbull1999}
\begin{eqnarray}
&r(s,t)= 1+ \epsilon\, \xi(s)\, cos(t+\alpha sin (t))\,,\nn \\
&z(s,t)= \kappa \epsilon\, \xi(s)\, sin (t)\,,\nn \\
&0\leq t\leq 2\pi\,, \; 0<s\leq 1\,, \label{D_domain}
\end{eqnarray}
where $\kappa$ is the vertical elongation of the torus,  $\epsilon=a_0/R_0$ is the inverse aspect ratio ($a_0$ and  $R_0$ are the  minor and major radius of the torus, respectively), and $\alpha:=arcsin(\delta)$, where $\delta$ is the triangularity of the boundary. In \eqref{D_domain} $\xi(s)$ is a monotonic function with the property $\xi(0)=0$ and $\xi(1)=1$. In this work we use $\xi(s)=s^k$, where $0<k<1$.

We create random point coordinates through Eq. \eqref{D_domain} within the domain $\Dc$ by generating random samples for the parameters $s$ and $t$ from uniform distributions over $[0, 1)$ and $[0,2\pi)$, respectively. This results in a random  distribution of points dispersed across the computational domain. Of course, one may create radially and poloidally uniform point distributions using Eq. \eqref{D_domain}, but we prefer to use random sets so as to avoid any geometric bias in the training process. With the parameter $k$ in $\xi(s)=s^k$ we adjust the radial density of the data points. For $k=0.5$ we have a nearly uniform density, while for $k<0.5$ and $k>0.5$ the density is higher near the boundary and near the center of the parametric domain, respectively. For the boundary points we set $s=1$ and let $t$ take either uniformly increasing or random values in $[0,2\pi)$. In Sec. \ref{sec_IV} we present several equilibria for up-down symmetric and asymmetric (diverted with lower X-point) configurations with ITER-relevant geometric characteristics, e.g., $\epsilon=2/6.2$, $\delta=0.4$ and $\kappa=1.6$. The sampling method for the diverted configurations with lower X-point is described in \ref{IV_D}.

\subsection{Equilibrium Parametrization}
\label{III_C}
Equation \eqref{ggse} contains three free functions that are related to the poloidal current density, the static pressure and the radial electric field. Here, these free functions are represented as truncated series expansions, i.e.
\begin{eqnarray}
  \frac{1}{2}\frac{X^2}{1-M_p^2} &=& X_0+\sum_{n=1}^{T} \frac{1}{n} X_{n} u^{n} \,,\nn\\
  P_s &=& P_0+\sum_{n=1}^{T} \frac{1}{n} P_n u^n\,\nn\\
  \frac{1}{2}\rho\left(\Phi'\right) &=& G_0+\sum_{n=1}^{T} \frac{1}{n} G_n u^n\,, \label{GGSE_Ansatz}
\end{eqnarray}
where $X_n,\, P_n,\, G_n$, $n=0,...,T$, are free parameters. In the following section we examine the cases  $T=1$, $T=2$ and $T=3$, adjusting the free parameters so as to obtain realistic values of the various physical quantities of interest but not necessarily realistic profiles. We aim to investigate the possibility to approximate realistic profiles upon inferring the values of the free parameters using experimental profile data. 

\medskip
\section{Results}
\label{sec_IV}
\subsection{Benchmarking against linear analytic solutions}
\label{IV_A}

\subsubsection{Solovev ansatz}
For $T=1$ in \eqref{GGSE_Ansatz} we obtain the so-called Solovev  ansatz \cite{Solovev1968}. In this case the homogeneous part of the GGSE residual consists of the Shafranov differential operator acting on $u$, i.e., $\Delta^*u$, and the inhomogeneous part is  $X_1+P_1r^2+G_1r^4$. In a previous work \cite{Kaltsas2014} we have derived an analytic solution to the Solovev-linearized GGSE and constructed tokamak equilibria with D-shaped cross sections with and without a lower X-point. This analytic solution is given by
\begin{eqnarray}
 u^*(r,z)=r\sum_{j}\Big[ a_j J_1(jr)e^{jz}+b_j J_1(jr)e^{-jz}\nn\\
 +c_jY_1(jr)e^{jz}+d_jY_1(jr)e^{-jz}\Big]\nn\\
 -\frac{X_1}{2}z^2-\frac{P_1}{8}r^4-\frac{G_1}{24}r^6\,,\label{solovev_solution}
\end{eqnarray}
where $J_1$ and $Y_1$ are the first-order Bessel functions of first and second kind, respectively and $a_j,b_j,c_j,d_j$ are arbitrary constants that are determined according to the desired boundary shaping.

\begin{figure}[!]
    \centering
    \includegraphics[scale=0.5]{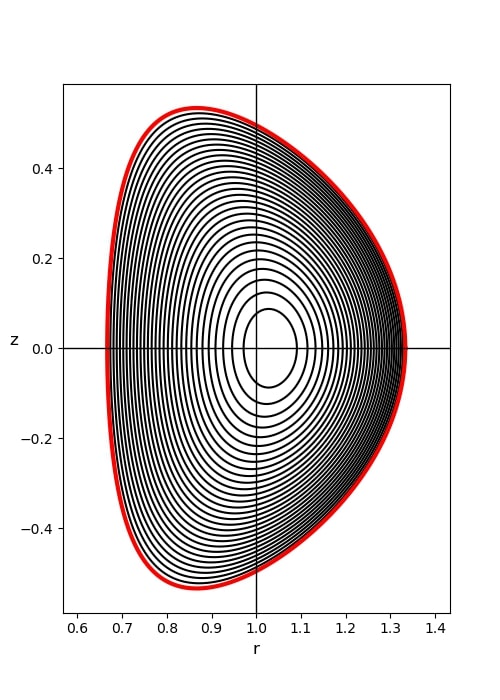}
    \includegraphics[scale = 0.38]{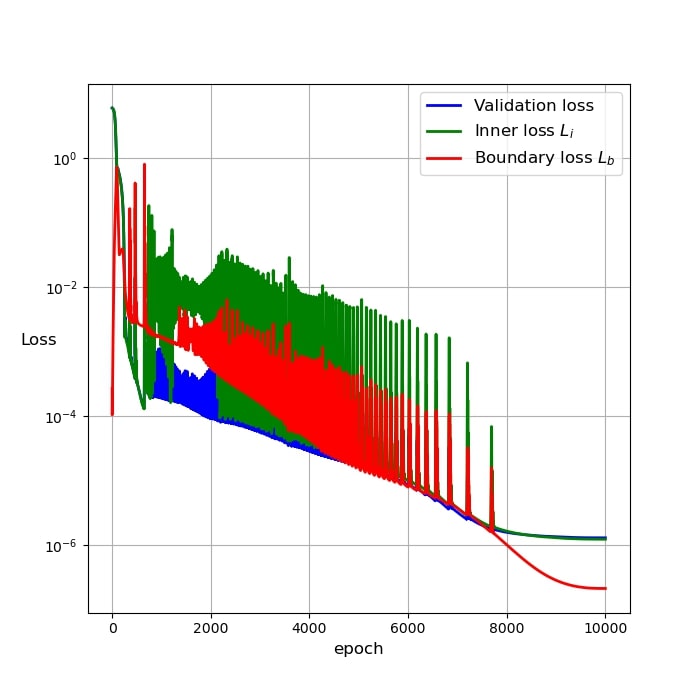}
    \caption{Magnetic configuration (top) and training history (bottom) of the Solovev Neural Network solution}
    \label{fig2_sol}
\end{figure}

We utilized this analytic solution to benchmark the accuracy of the resulting neural network solution. The training was performed on 1024 randomly allocated points in the domain $\Dc$ and 1024 points on the D-Shaped boundary $\partial \Dc$. The poloidal cross section of the magnetic surfaces is shown in Fig. \ref{fig2_sol} along with the training history of the corresponding neural network solution. A loss function computed on a set of inner validation points has been incorporated in the same plot. It can be seen that the validation loss is lower than the training loss (green line), hence there is no overfitting on the inner training points. 

\begin{figure}[h!]
    \centering
    \includegraphics[scale = 0.4]{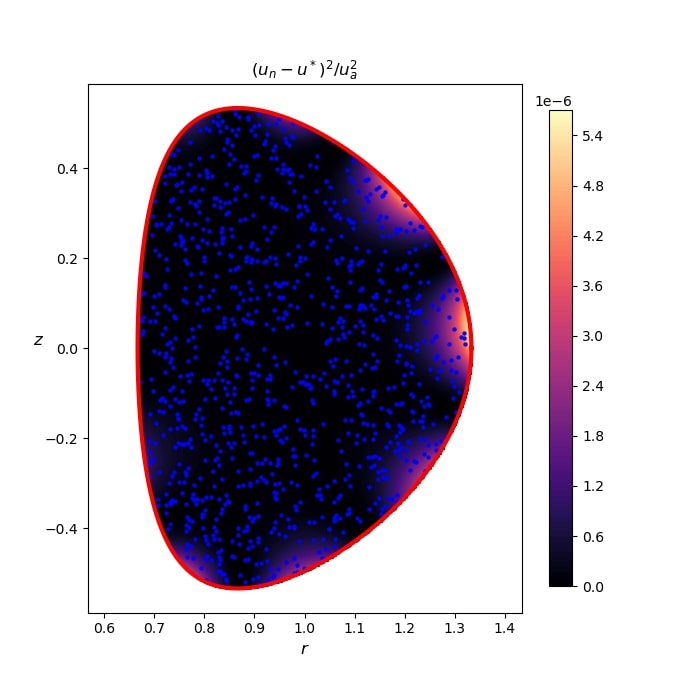}
    \includegraphics[scale = 0.5]{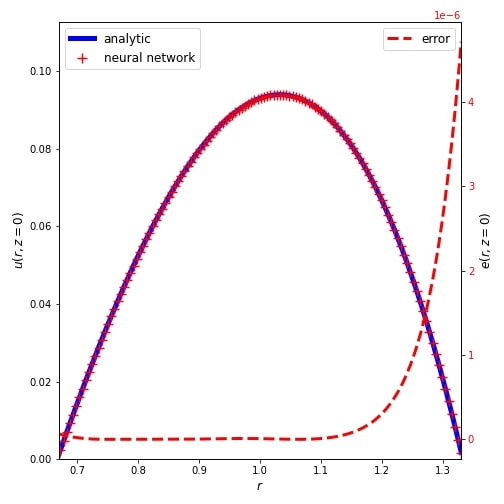}
    \caption{Top: estimated error of the Solovev neural network solution compared with the analytic solution \eqref{solovev_solution}. The distribution of the randomly sampled training points (blue dots) is also presented. Bottom: comparison of the Solovev neural network with the analytic solution \eqref{solovev_solution} on the equatorial plane $z=0$.}
    \label{fig3_sol}
\end{figure}

In Fig. \ref{fig3_sol} we compare the neural network solution with the solution \eqref{solovev_solution}. The relative error, which is defined by
\begin{eqnarray}
e = \frac{|u_n-u^*|^2}{|u_{a}|^2}\,,
\end{eqnarray}
where $u_a$ is the value of $u_n$ on the magnetic axis, is of the order of $10^{-6}$ after $10000$ epochs. In Fig. \ref{fig3_sol} we observe that the error is rather insignificant in the greater part of the cross section, while it attains larger values in certain regions near the boundary. A comparison with the analytic solution on the equatorial ($z=0$) plane is also presented confirming the remarkable accuracy of the neural network solution. We stress that comparisons for various values of the free parameters were performed verifying the good agreement between the neural network  and the analytic solutions.

\subsubsection{General linearizing ansatz}
For the general linearizing ansatz, given by \eqref{GGSE_Ansatz} with $T=2$, the GGSE residual is given by
\begin{eqnarray}
 \Rc=\partial_{rr}u-(\partial_r u)/r+\partial_{zz} u +X_1+P_1 r^2 +G_1 r^4\nn\\
 +(X_2+P_2 r^2 +G_2 r^4)u\,. \label{linear_residual}
\end{eqnarray}
For $G_1=G_2=0$ analytic solutions to $\Rc=0$ have been found in Ref. \onlinecite{Atanasiu2004}, while for the more general case $G_1,G_2\neq 0$, we derived in Ref. \onlinecite{Kaltsas2019c}  a general analytic solution of the form:
\begin{eqnarray}
 u^*(r,z)=\sum_{n=0}^{K_1}f_n(r)z^n+\sum_{n=0}^{K_2}a_n(r-1)^n\,,  \label{analytic}
\end{eqnarray}
where $f_n(r)$ are determined using the Frobenius method applied for the regular singular point $r=0$. We showed that, keeping sufficiently large numbers of terms, when truncating the various power series involved in this analytic solution, the GGSE is satisfied up to machine precision (see Ref. \onlinecite{Kaltsas2019c} for details). 

\begin{figure}[h!]
    \centering
    \includegraphics[scale = 0.4]{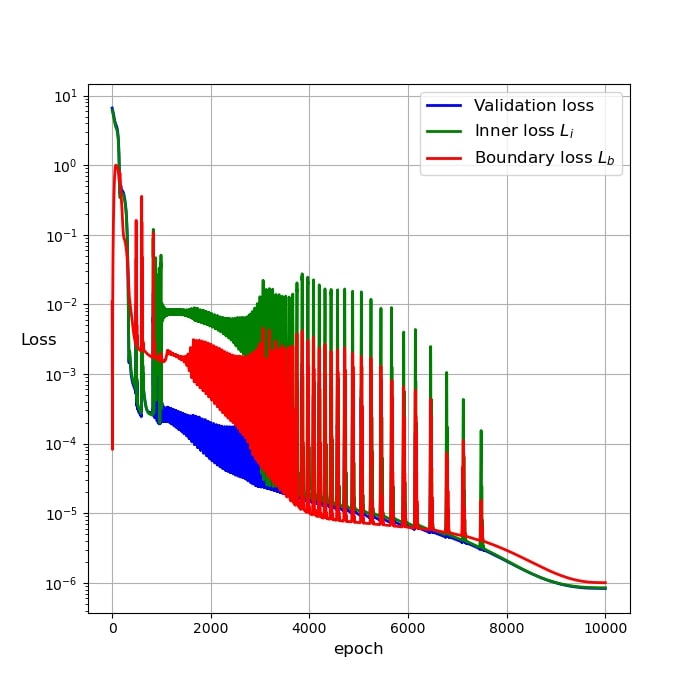}
    \caption{ Training history for the linear neural network equilibrium solution.}
    \label{fig1_lin}
\end{figure}

\begin{figure}[h!]
    \centering
    \includegraphics[scale = 0.4]{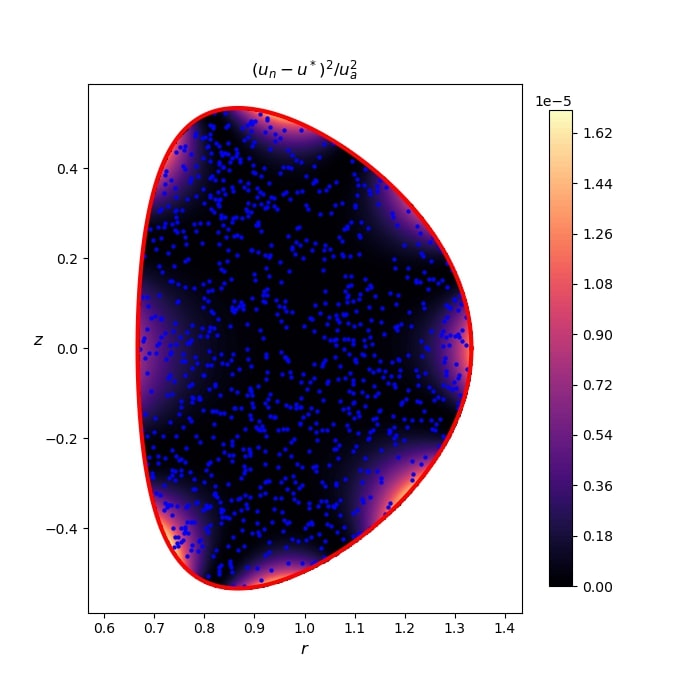}
        \includegraphics[scale = 0.5]{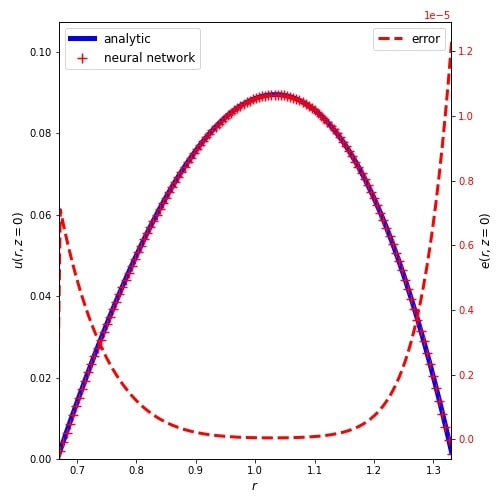}
    \caption{Top: estimated error between the neural network and the analytic solution of the form \eqref{analytic} in conjuction with the randomly sampled training points (blue points). Bottom: comparison of the linear neural network and the analytic solution on the equatorial plane $z=0$.}
    \label{fig3_lin}
\end{figure}

Here, we compute a linear neural network equilibrium upon minimizing the losses \eqref{loss_inner}, \eqref{loss_boundary}. The geometric and profile parametric values that have been used correspond to the particular analytic equilibrium of Ref. \onlinecite{Kaltsas2019c}. The training history of the network is shown in Fig. \ref{fig1_lin}, while the resulting magnetic configuration is similar to the Solovev equilibrium. In addition, the neural network solution is benchmarked against the analytic solution Eq. \eqref{analytic} (Fig. \ref{fig3_lin}) as done previously for the Solovev solution. The high degree of agreement between the two solutions was confirmed for several choices of free parameters.

\subsection{Interpolation capabilities of the neural network solutions}
\label{IV_B}
The training point sampling method described in \ref{III_B} allows us to adjust the radial distribution of the inner collocation points by adjusting the parameter $k$ in Eq. \eqref{D_domain}.  Varying $k$, we can investigate the impact of the radial distribution of points on the training process and on the accuracy of the neural network solution and assess its interpolation capabilities in extended regions with no collocation points.

\begin{figure}[h!]
    \centering
    \includegraphics[scale=0.45]{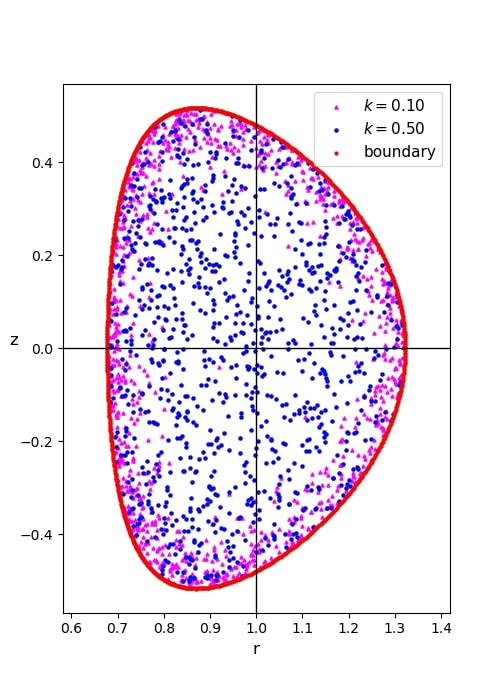}
    \caption{Distribution of training points for $k=0.50$ (blue dots) and $k=0.10$ (magenta triangles). The boundary points (red dots) are uniformly distributed in both cases.}
    \label{fig1_inter}
\end{figure}

\begin{figure}[!]
    \centering
    \includegraphics[scale=0.5]{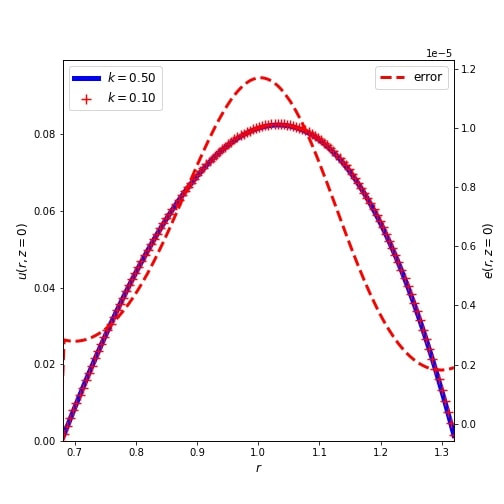}\\
    \includegraphics[scale=0.4]{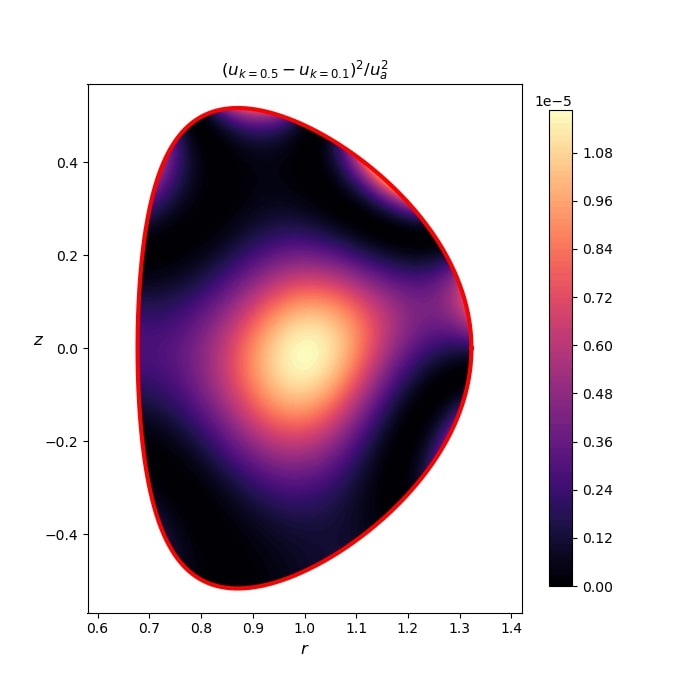}
    \caption{Top: comparison between two neural network solutions trained on points with different radial distributions. Bottom: the relative error of the two solutions. }
    \label{fig3_inter}
\end{figure}

Here we compare a neural network trained on collocation points generated by \eqref{D_domain} with $k=0.10$, with the neural network solution of the previous subsection where $k=0.50$. The two training sets are superimposed in Fig. \ref{fig1_inter}; a comparison between the two solutions  and an estimation of the relative error is seen in Fig. \ref{fig3_inter}. The maximum relative error is of the order of $10^{-5}$ after 10000 epochs. This result indicates that the neural network solutions can accurately describe plasma equilibria all over the computational domain even if the training is performed at its edge region. 

\subsection{Nonlinear Equilibrium with H-mode characteristics}
A nonlinear GGSE is obtained for  $T>2$. Although we present only the case $T=3$, we have observed that the MLP algorithm performs well also for $T>3$,  depending, however, on the relative strength of the nonlinear terms. For strong nonlinear contributions the algorithm should be initialized with appropriately pre-trained networks to ensure convergence.

In the particular case $T=3$ we have:
\begin{eqnarray}
 \Rc=\partial_{rr}u-(\partial_r u)/r+\partial_{zz} u +X_1+P_1 r^2 +G_1 r^4\nn\\
 +(X_2+P_2 r^2 +G_2 r^4)u + (X_3+P_3 r^2 +G_3 r^4)u^2\,.
\end{eqnarray}

\subsubsection{Constrained minimization}
\begin{figure}[!]
    \centering
    \includegraphics[scale=0.5]{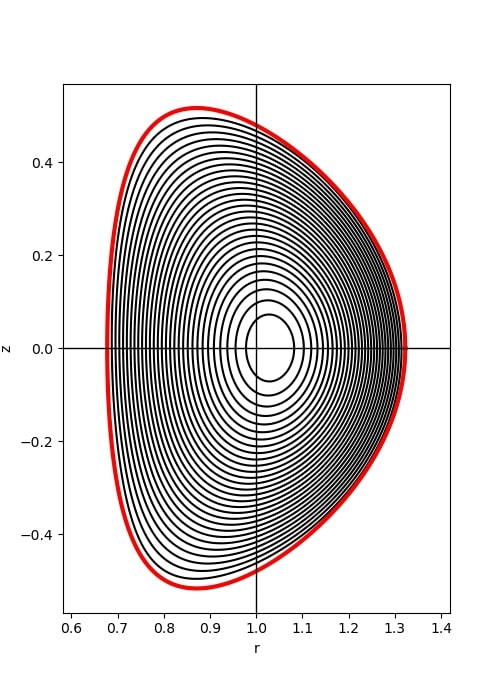}\\
    \includegraphics[scale=0.38]{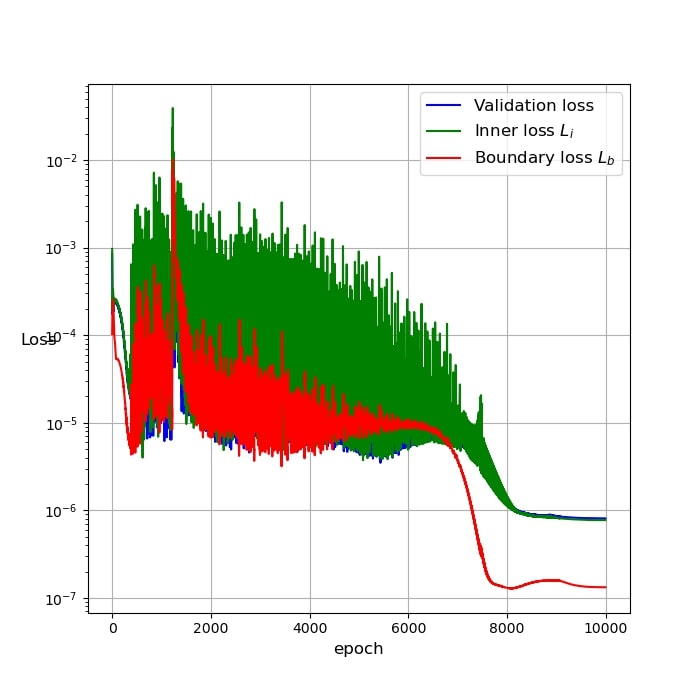}
    \caption{The magnetic field configuration for the nonlinear ansatz $(T=3)$ (top) and the training history of the corresponding neural network solution (bottom).}
    \label{fig_7}
\end{figure}

To facilitate the parameter tuning we let $P_2$ and $X_2$ to be learned by the training process upon requiring the average $\beta$ and the total toroidal current to take specific, predetermined values, namely
\begin{eqnarray}
 \langle \beta \rangle&=&0.03\nn\\
 I_t&=&15\, MA\,,\nn
\end{eqnarray}
These constraints are imposed by incorporating the following penalty terms in the total loss function
\begin{eqnarray}
 &&L_{\langle\beta\rangle}=\bigg|\frac{1}{N}\sum_{j=1}^{N} P(r_j,u_n(r_j,z_j))-\langle \beta\rangle \bigg|^2\,, \label{loss_beta}\\
 &&\hspace{-3mm } L_{I}= \bigg|\frac{S}{N}\sum_{j=1}^{N} J_\phi(r_j,u_n(r_j,z_j))-I_t \frac{\mu_0}{R_0B_0}\bigg|^2\,, \label{loss_I}
\end{eqnarray}
where, $S$ is the poloidal cross-sectional area, $P$ is the Alfv\'en-normalized dynamic pressure and
\begin{eqnarray}
 \hspace{-5mm}J_\phi=\frac{1}{r [1-M_p^2(u)]^{1/2}}\left(\Delta^*u+\frac{M_p(u)M_p'(u)}{1-M_p^2(u)}|\nb u|^2\right)\,,
\end{eqnarray}
is the normalized toroidal current density.  The resulting magnetic configuration and training history are shown in Fig. \ref{fig_7}.

\subsubsection{H-mode equilibrium}
 The mechanisms behind the L-H transition and the associated phenomenology, e.g., the transport barrier and density pedestal formation, is not yet fully understood even from the force balance point of view, although since its discovery in 1982 \citep{Wagner2007}, most tokamaks are designed to operate in H-mode. Therefore, the construction of equilibrium states encompassing H-mode characteristics such as density and pressure pedestals (see, for example, the recent works  \onlinecite{Kaltsas2019c,Li2021,Guazzotto2021b,Montani2021}) is important for fusion research. Moreover, the construction of such states might be useful for performing plasma stability and transport studies concerning the H-mode operational regime. 

\begin{figure}[h!]
    \centering
    \includegraphics[scale=0.55]{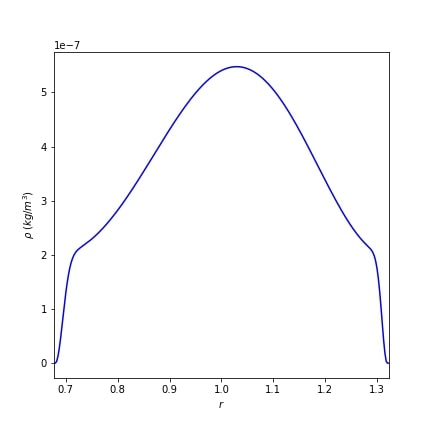}\\
    \includegraphics[scale=0.55]{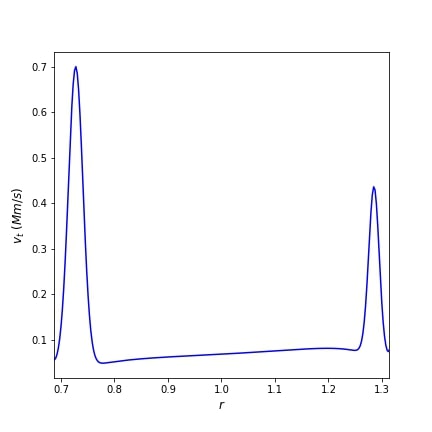}\\
    \caption{The mass density (top), and the toroidal velocity profile (bottom) plotted on the midplane $z=0$.}
    \label{fig1_H}
\end{figure}

\begin{figure}[h!]
    \centering
    \includegraphics[scale=0.55]{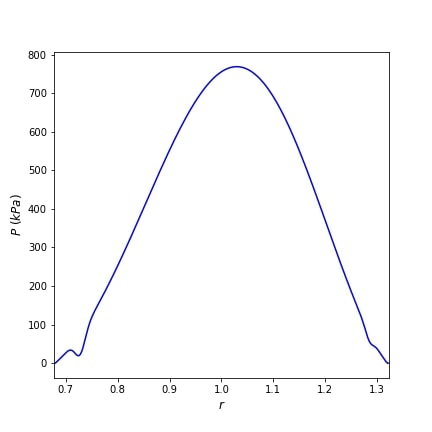}\\
    \includegraphics[scale=0.55]{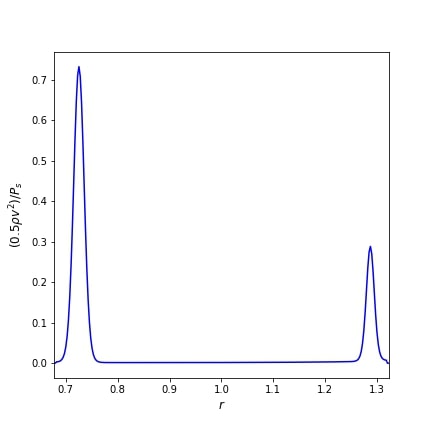}
    \caption{The total pressure profile (top) and the rotational to kinetic energy density ratio (bottom), plotted against the cylindrical coordinate $r$. The rotational to kinetic energy ratio is $0.02$.}
    \label{fig1.5_H}
\end{figure}

\begin{figure}[!]
    \centering
    \includegraphics[scale=0.55]{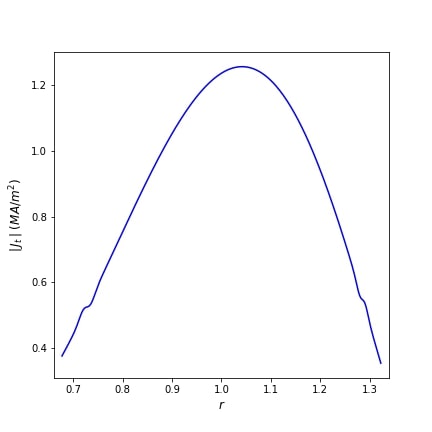}\\
    \includegraphics[scale=0.55]{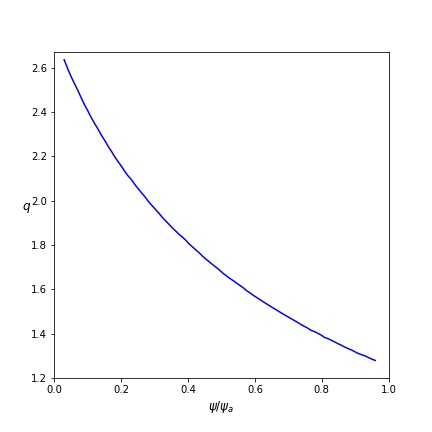}
    \caption{The toroidal current density profile (top) plotted on the midplane $z=0$ and the safety factor (bottom) plotted against  $\psi/\psi_a$, from $\psi/\psi_a=0.03$ to $0.97$. Our definition of poloidal flux increases from $\psi/\psi_a=0$ at the boundary to $\psi/\psi_a=1$ at the magnetic axis.}
    \label{fig2_H}
\end{figure}

An H-mode-pertinent ansatz \cite{Pataki2013} that results in a mass density pedestal is given by
\begin{eqnarray}
 \rho(u)=(\rho_1+\rho_2u^2)\left(1-\exp{\left(-u^2/\mu\right)}\right)\,,
\end{eqnarray}
while for the Mach function $M_p(u)$ we choose a shifted Gaussian \cite{Li2021}, i.e.
\begin{eqnarray}
 M_p(u)=M_0\exp{\left(-\frac{(u-u_f)^2}{2\sigma^2}\right)}\,,
\end{eqnarray}
where $M_0$ is the maximum Mach number, $u_f$ is the value of $u$ on the magnetic surface where $M_p$ is maximum and $\sigma$ is the width of the Gaussian profile. Using these free functions and the neural network solutions for the  nonlinear GGSE we define a tokamak equilibrium encompassing some H-mode characteristics. In Fig. \ref{fig1_H} we present the mass density profile that exhibits a characteristic pedestal morphology and the computed toroidal velocity profile. In Fig. \ref{fig1.5_H} the pressure and the rotational to kinetic energy density ratio profiles are provided while in Fig. \ref{fig2_H} we present the toroidal current density and the safety factor profile.  Using the general linear solution we obtained similar profiles (not presented here) with noticeable but not major differences. Significant differences can be observed upon increasing the parameters $P_3$ and $X_3$. In this case though, the convergence deteriorates if the constraints \eqref{loss_beta} and \eqref{loss_I} are preserved. 

The toroidal velocity is given by the following expression
\begin{eqnarray}
v_\phi= \frac{M_p}{r\sqrt{\rho}}I-r\Phi'(u)\sqrt{1-M_p^2}\,,
\end{eqnarray}
while the plasma pressure can be computed by a Bernoulli equation \cite{Tasso1998}
\begin{eqnarray}
P=P_s(u)-\rho \left\{\frac{v^2}{2}-r^2\left[\Phi'(u)\right]^2\right\}\,,\label{Bernoulli}
\end{eqnarray}
where $v^2=\sqrt{v_{\phi}^2+v_{p}^2}$, with $$v_{p}^2=\left(\frac{M_p}{r\sqrt{\rho}}\right)^2|\nb u|^2\,.$$

\subsection{Equilibrium with lower X-point}
\label{IV_D}
For a diverted configuration with a lower X-point in the separatrix, the following parametric equations \citep{Kuiroukidis2015} are used to define the computational domain:
\begin{eqnarray}
r(s,t) &=&
\begin{cases}
1+ \epsilon\, \xi(s)\, cos(t+\alpha sin (t))\,,& 0\leq t\leq \pi\,,\nn \\
1+\epsilon \xi(s) cos(t)\,,& \pi\leq t \leq 2\pi\,,
\end{cases}\\\label{DX_domain}
\\
z(s,t) &=&
 \begin{cases}
\kappa \epsilon\, \xi(s)\, sin (t)\,, &
0\leq t\leq \pi\,, \nn\\
-\kappa_d\epsilon\left[\frac{\xi(s)(1+cos(t))}{1+cos(\theta_d)}\right]^{1/2}\,, & \pi\leq t< 2\pi-\theta_d\,,\nn\\
-\kappa_d\epsilon\left[\frac{\xi(s)(1-cos(t))}{1+cos(\theta_d)}\right]^{1/2}\,, & 2\pi-\theta_d\leq t\leq 2\pi\,,
\end{cases}
\end{eqnarray}
where $ 0 \leq s\leq 1$ and $ \theta_d=\pi-arctan(\kappa_d/\delta_d)$.

\begin{figure}[!]
    \includegraphics[scale=0.5]{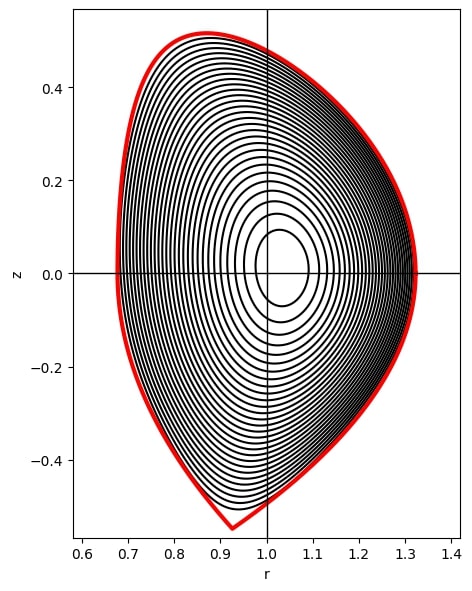}
    \includegraphics[scale=0.4]{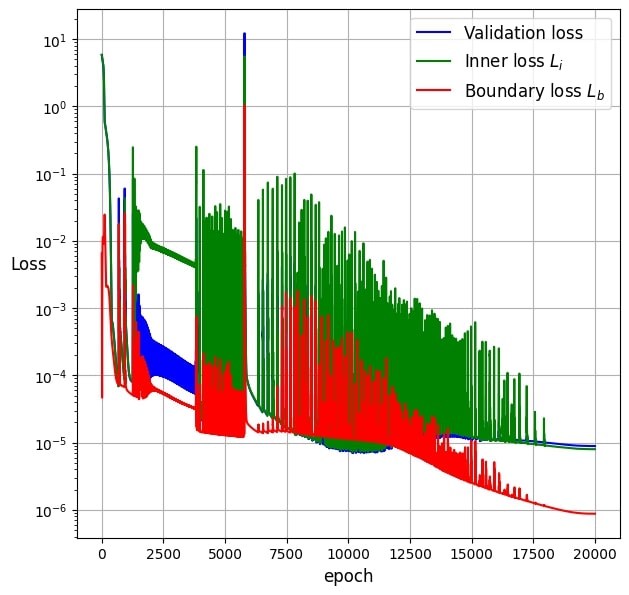}
    \caption{Equilibrium configuration with lower X-point in connection with the neural network solution of the general linearized GGSE (top). The training history of the neural network equilibrium with lower X-point (bottom).}
    \label{fig_x-point}
\end{figure}

For the X-point equilibrium we had to double the number of collocation points and epochs in order to obtain a magnetic configuration with the desired geometric characteristics. Namely, $N_i=N_b=2048$ and the neural network has been trained for 20000 epochs. The resulting magnetic configuration and the corresponding training history of the neural network are presented in Fig. \ref{fig_x-point}. 

\medskip
\section{Conclusions}
\label{sec_V}
In this work we computed neural network solutions to a generalized Grad-Shafranov equation (GGSE), governing axisymmetric MHD equilibria with incompressible flows of arbitrary direction. The neural networks were trained in domains with tokamak-pertinent shapes adopting linear and nonlinear choices for the free functions which are involved in the GGSE. The solutions were benchmarked against exact, analytic solutions derived in previous works, displaying fairly good agreement.  In addition, the remarkable capability of the neural networks to approximate equilibrium solutions in regions with sparse or absent collocation points was demonstrated upon localizing the distribution of the training points near the plasma edge. The results presented here will be further corroborated upon benchmarking against tested numerical equilibrium solvers such as HELENA for realistic equilibrium profiles in a future work. Comparisons regarding the accuracy and the flexibility of the solutions in conjunction with the corresponding CPU time will also be attempted\footnote{To provide a quantitative measure of CPU time, we executed the code in a personal computer with an Intel i7-7500U CPU at 2.70GHz ($2$ cores). For the nonlinear example of Sec. \ref{sec_IV} with 2048 training points and $12737+3$ learnable parameters, each epoch takes approximately $0.1$ $s$, while the entire runtime is approximately $17$ minutes}.\\
\section*{Acknowledgements}
This work has  received funding from  the National Programme for the Controlled Thermonuclear Fusion, Hellenic Republic.  The authors would like to thank George Poulipoulis for fruitful discussions on plasma equilibrium. D.A.K. is also thankful to Ioakim Chorozidis for interesting debates on Python related topics.
\section*{Author Declarations}
\subsection*{Conflict of Interest}
The authors declare that there is no conflict of interest.
%
\section*{Data availability}
The data that support the findings of this study are available from the corresponding author upon reasonable request.
%
%
\section*{References} 
\bibliography{biblio.bib}

\end{document}